\documentclass{article}
\usepackage{spconf,amsmath,graphicx}
\usepackage{cite}
\usepackage{amsmath,amssymb,amsfonts}
\usepackage{algorithmic}
\usepackage{graphicx}
\usepackage{textcomp}
\usepackage{xcolor}
\usepackage{url}

\newcommand{\mysize}{2.05}
\usepackage[tight,footnotesize]{subfigure}
\usepackage{tablefootnote}
\usepackage{booktabs}
\usepackage{diagbox}

\title{Speech Emotion Recognition with Multiscale Area Attention and Data Augmentation}
\name{Mingke Xu$^1$, Fan Zhang$^2$, Xiaodong Cui$^3$, Wei Zhang$^3$}
\address{$^1$Nanjing Tech University, China  \quad$^2$IBM Data and AI, USA  \quad$^3$IBM Research AI, USA }

\begin{document}
%
\maketitle

\begin{abstract}


In Speech Emotion Recognition (SER), emotional characteristics often appear in diverse forms of energy patterns in spectrograms. Typical attention neural network classifiers of SER are usually optimized on a fixed attention granularity. In this paper, we apply multiscale area attention in a deep convolutional neural network to attend emotional characteristics with varied granularities and therefore the classifier can benefit from an ensemble of attentions with different scales. To deal with data sparsity, we conduct data augmentation with vocal tract length perturbation (VTLP) to improve the generalization capability of the classifier. Experiments are carried out on the Interactive Emotional Dyadic Motion Capture (IEMOCAP) dataset. We achieved 79.34\% weighted accuracy (WA) and 77.54\% unweighted accuracy (UA), which, to the best of our knowledge, is the state of the art on this dataset.
\end{abstract}

\noindent\textbf{Index Terms}: speech emotion recognition, convolutional neural network, attention mechanism, data augmentation

\section{Introduction}

Speech is an important carrier of emotions in human communication. Speech Emotion Recognition (SER) has wide application perspectives on psychological assessment\cite{low2010detection}, robots\cite{huahu2010application}, mobile services\cite{yoon2007study}, \emph{etc}. 
For example, a psychologist can design a treatment plan according to the emotions hidden/expressed in the patient’s speech. Deep learning has accelerated the progress of recognizing human emotions from speech \cite{han2014speech,chen20183,wu2019speech,xu2020hgfm,priyasad2020attention,nediyanchath2020multi}, but there are still deficiencies in the research of SER, such as data shortage and insufficient model accuracy. 

Recently, we proposed Head Fusion Net \cite{xu2020improve}\footnote{The code is released at github.com/lessonxmk/head\_fusion} which achieved the state-of-the-art performance on the IEMOCAP dataset. However, it does not fully address the above problems. In SER, emotion may display distinct energy patterns in spectrograms with varied granularity of areas. However, typical attention models in SER are usually optimized on a fixed scale, which may limit the model's capability to deal with diverse areas and granularities. Therefore, in this paper, we introduce multiscale area attention to a deep convolutional neural network model based on Head Fusion to improve model accuracy. Furthermore, data augmentation is used to address the data scarcity issue.


Our main contributions are as follows:
\vspace{-6 pt}
\begin{itemize}
\setlength{\itemsep}{0pt}
\setlength{\parsep}{0pt}
\setlength{\parskip}{0pt}
    \item To the best of our knowledge, this is the first attempt for applying multiscale area attention to SER. 
    \item We performed data augmentation on the IEMOCAP dataset with vocal tract length perturbation (VTLP) and achieved an accuracy improvement of about 0.5\% absolute.
    \item With area attention and VTLP-based data augmentation, we achieved the state-of-the-art on the IEMOCAP dataset with an WA of 79.34\% and UA of 77.54\%.\footnote{The code is released at github.com/lessonxmk/Optimized\_attention\_for\_SER} 
\end{itemize}

\section{Related work}
In 2014, the first SER model based on deep learning was proposed by Han \emph{et al.}\cite{han2014speech}.  Recently, for the same purpose, M. Chen \emph{et al.}\cite{chen20183} combined convolutional neural networks (CNN) and Long Short-Term Memory (LSTM); X. Wu \emph{et al.}\cite{wu2019speech} replaced CNN with capsule networks (CapsNet); Y. Xu \emph{et al.}\cite{xu2020hgfm} used Gate Recurrent Unit (GRU) to calculate features from frame and utterance level, and S. Parthasarathy\cite{parthasarathy2020semi} used ladder networks to combine the unsupervised auxiliary task and the primary task of predicting emotional attributes.

There is a recent resurgence of interest on attention-based SER models \cite{tarantino2019self,priyasad2020attention,nediyanchath2020multi}. 
However, those attention mechanisms can only be calculated with a preset granularity which may not adapt dynamically to different areas of interest in spectrogram. Y. Li \emph{et al.} \cite{li2019area} proposed area attention that allows the model to calculate attention with multiple granularities concurrently, an idea that is not yet explored in SER.

Insufficient data hinders progress in SER. Data augmentation has become a popular method to increase training data\cite{jaitly2013vocal,cui2015dataconv,cui2015data,park2019specaugment} in the related field of Automatic Speech Recognition (ASR). Yet, it has not enjoyed broad attention for SER.

In this paper, we extend the multiscale area attention to SER with data augmentation. We introduce our method in section \ref{sec:method} and experiment results in section \ref{sec:exp} followed by the conclusion with section \ref{sec:conclusion}.
\begin{figure}[htbp]
    \centering
    \includegraphics[width=0.8\columnwidth]{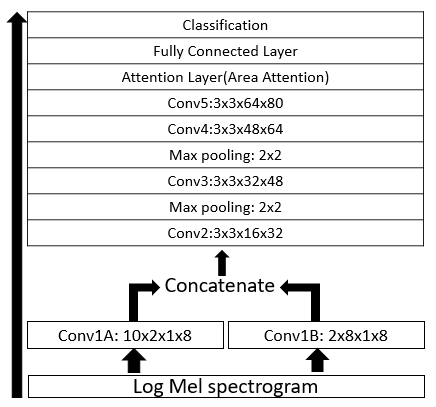}
    \caption{The architecture of the CNN with attention used as a classifier in this work.  
    }
    \label{model}
\end{figure}
\section{Methodology}
\label{sec:method}
We first introduce our base convolutional neural networks that shares similarity to our Head Fusion Net \cite{xu2020improve}, then our newly introduced multiscale area attention that enhances Head Fusion Net, and finally the data augmentation technique. 
\subsection{Convolutional Neural Networks}
We designed an attention-based convolutional neural network with 5 convolutional layers, an attention layer, and a fully connected layer. Fig. \ref{model} shows the detailed model structure. First, the Librosa audio processing library\cite{2015librosa} is used to extract the logMel spectrogram as features, which are fed into two parallel convolutional layers to extract textures from the time axis and frequency axis, respectively. The result is fed into four consecutive convolutional layers and generates an 80-channel representation. Then the attention layer attends on the representation and sends the outputs to the fully connected layer for classification. Batch normalization is applied after each convolutional layer.
\vspace{-10 pt}

\subsection{Multiscale Area Attention}


In this section, We extend the area attention in Y. Li \emph{et al.} \cite{li2019area} to SER. The attention mechanism can be regarded as a soft addressing operation, which uses key-value pairs to represent the content stored in the memory, and the elements are composed of the address (\emph{key}) and the value (\emph{value}). \emph{Query} can match to a key which correspondent \emph{value} is retrieved from the memory according to the degree of correlation between the \emph{query} and the \emph{key}. The \emph{query}, \emph{key}, and \emph{value} are usually first multiplied by a parameter matrix \emph{W} to obtain \emph{Q, K} and \emph{V}. Eq.\ref{attn} shows the calculation of attention score, where \emph{$d_k$} is the dimension of \emph{K}\cite{vaswani2017attention} to prevent the result from being too large.
\begin{equation}
    \begin{split}
    & Q=W_q*query, K=W_k*key, V=W_v*value \\
    & Attention(Q,K,V)=softmax\left(\frac{QK^T}{\sqrt{d_k}}\right)V \\
    \end{split}
    \label{attn}
\end{equation}
\vspace{-10pt}



In self-attention, the \emph{query}, \emph{key} and \emph{value} come from a same input \emph{X}. By calculating self-attention, the model can focus on the connection between different parts of the input. In SER, the distribution of emotion characteristics often crosses a larger scale, and using self-attention in speech emotion recognition improves the accuracy.

However, under the conventional attention, the model only uses a preset granularity as the basic unit for calculation, \emph{e.g.}, a word for a word-level translation model, a grid cell for an image-based model, \emph{etc.} Yet, it is hard to know which granularity is most suitable for a complex task. 

\begin{figure}[htbp]
    \centering
    \includegraphics[width=0.83\columnwidth]{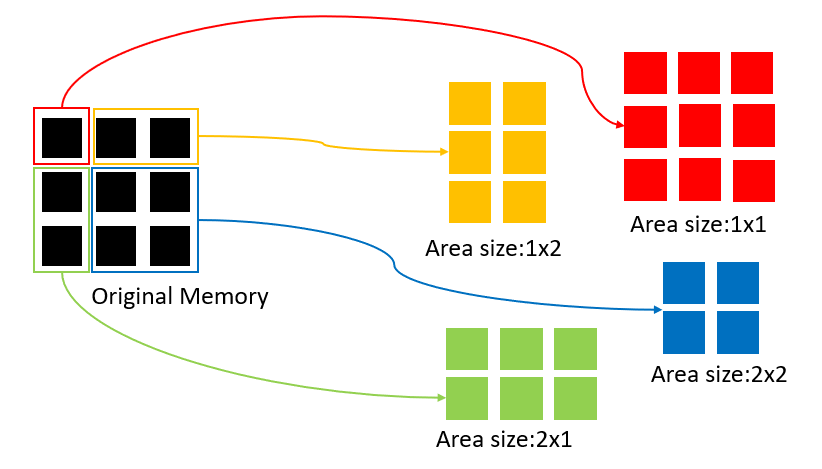}
    \caption{\textbf{Generating Area} Multiple areas can be generated by combining adjacent items in a continuous memory block. For a 3x3 memory block, if we set the max area size to 2x2, the memory block can be divided into 1x1,1x2,2x1 and 2x2.  }
    \label{area}
\end{figure}

Area attention allows the model to attend at multiple scales and granularities and to learn the most appropriate granularities. As shown in Fig. \ref{area}, for a continuous memory block, multiple areas can be created to accommodate for different granularities, \emph{e.g.}, 1x2, 2x1, 2x2 and \emph{etc.}. In order to calculate attention in units of areas, we need to define the \emph{key} and \emph{value} for the area. For example, we can define the mean of an area as the \emph{key} and the sum of an area as the \emph{value}, so that the attention can be evaluated in a way similar to ordinary attention. (Eq.\ref{attn})

Exhaustive evaluation of attention on a large memory block may be computationally prohibitive. A maximum length and width is set to an area under investigation.  


\vspace{-20pt}
\subsection{Data Augmentation}
Given the limited amount of training data in IEMOCAP,  we use vocal tract length perturbation (VTLP)\cite{jaitly2013vocal} as means for data augmentation. VTLP increases the number of speakers by perturbing the vocal tract length. We generated additional 7 replicas of the original data with nlpaug library\cite{ma2019nlpaug}. The augmented data is only used for training.

\begin{figure*}[htbp]
    \centering
    \subfigure[On the original data set]{\includegraphics[width=\mysize in]{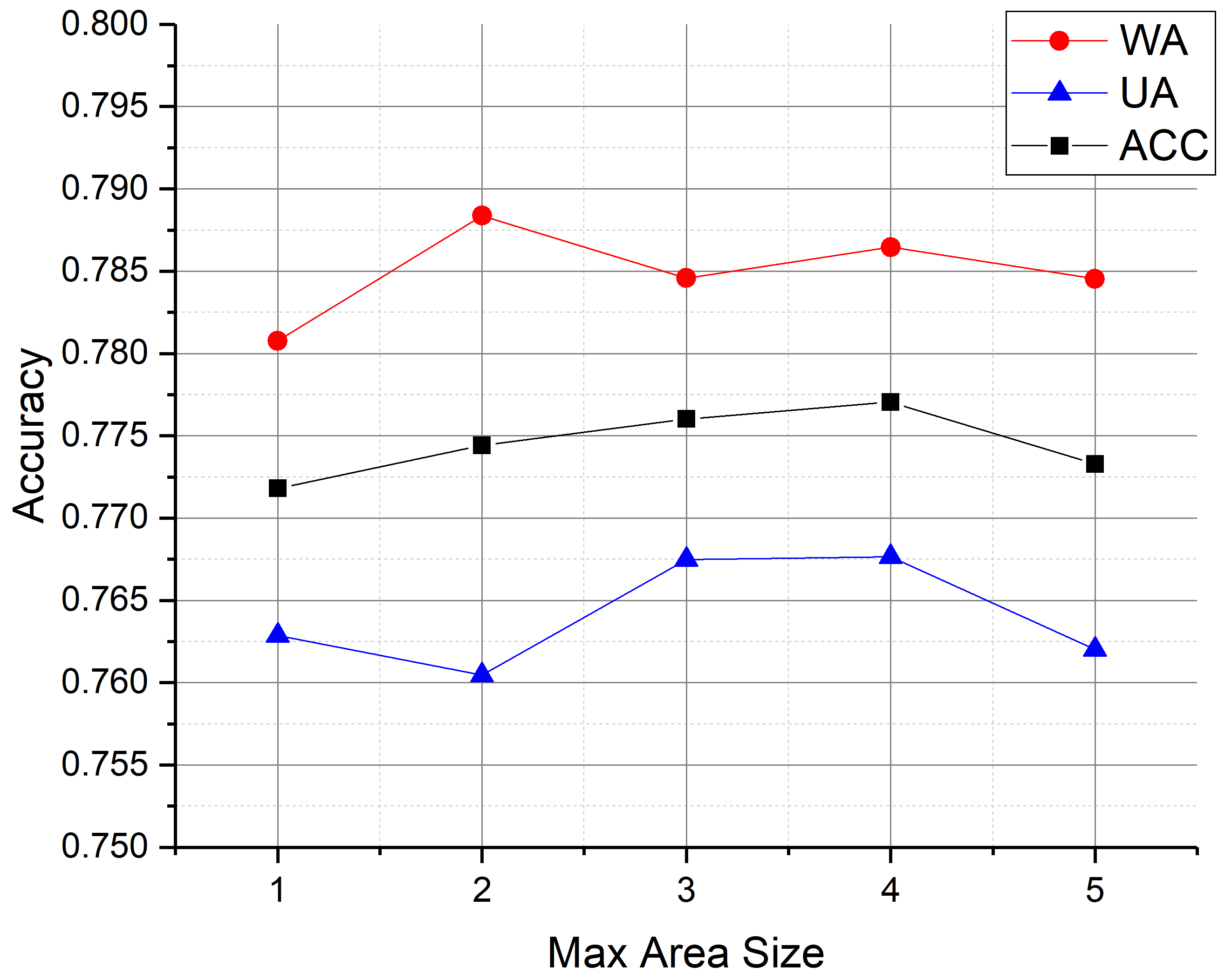}}
    \subfigure[On the augmented data set]{\includegraphics[width=\mysize in]{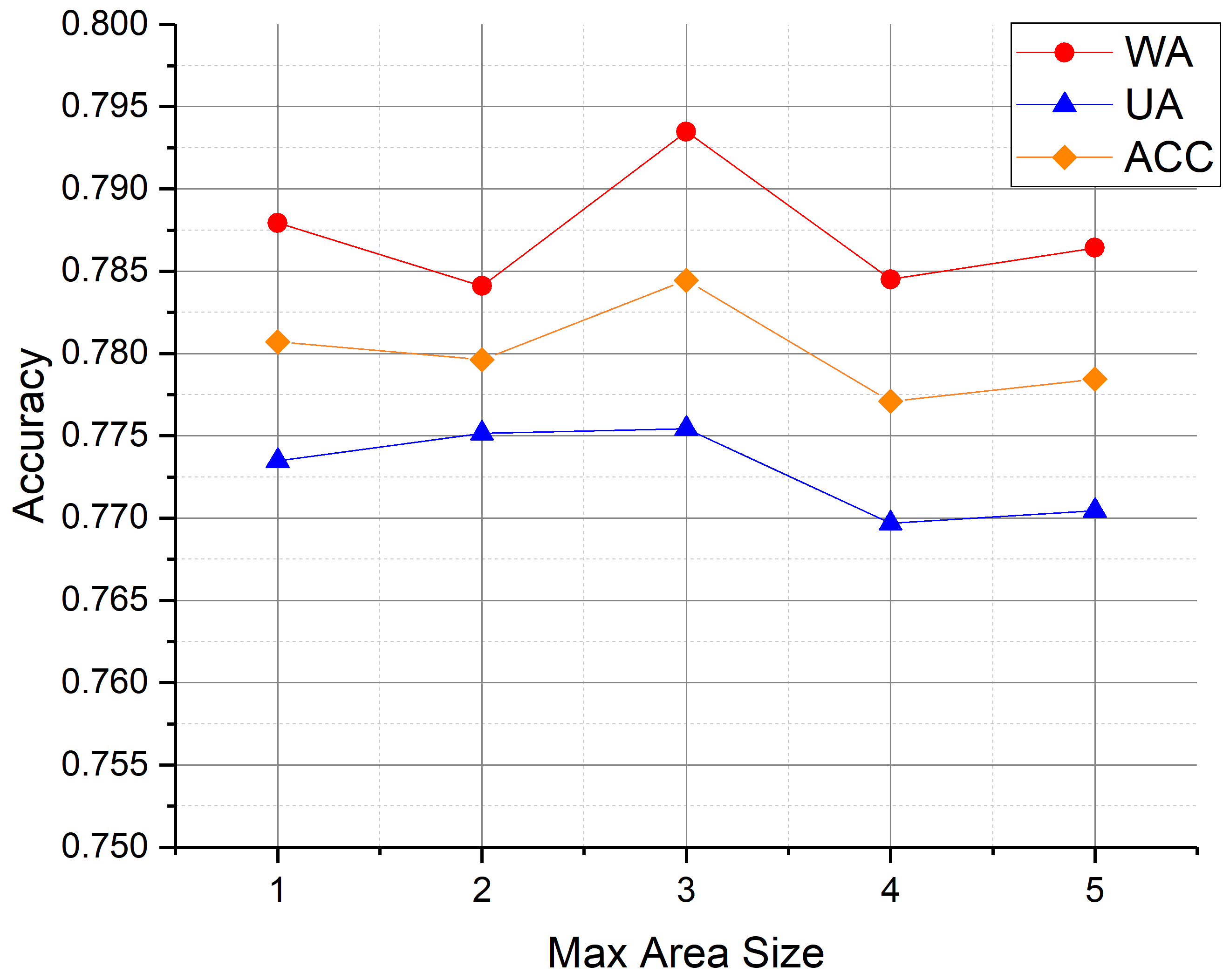}}
    \subfigure[comparison of \emph{ACC}]{\includegraphics[width=\mysize in]{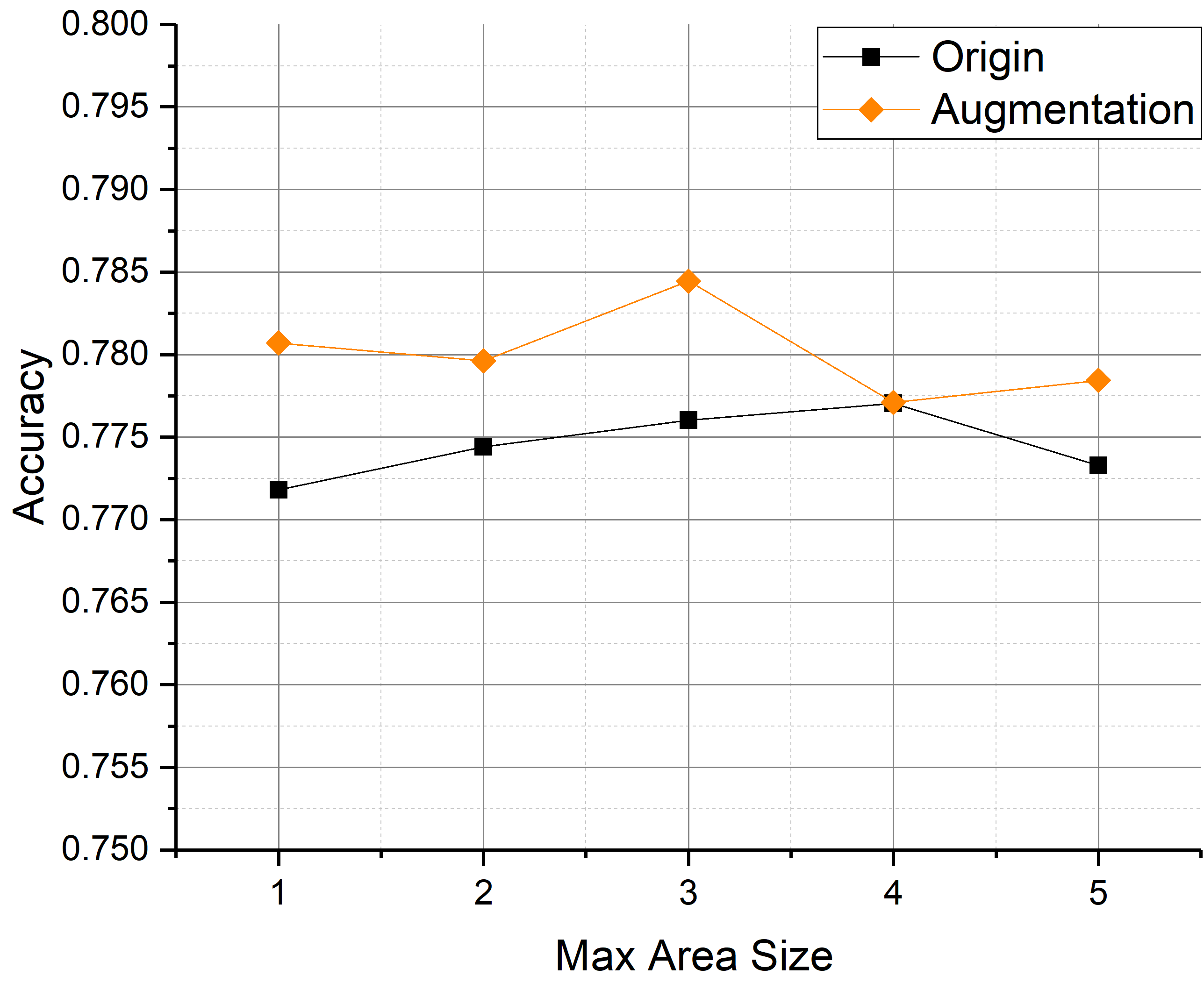}}
    \caption{\textbf{Result of modifying max area size} It can be seen that when trained on the original data set, the model with the max area size of 4x4 achieved the highest \emph{ACC}, followed by 3x3. When trained on the augmented data set, the model with the max area size of 3x3 achieved the highest \emph{ACC}. In most cases, the use of enhanced data brings an accuracy increase of more than 0.5\%. The best model achieved 78.44\% \emph{ACC}, where WA=79.34\% and UA=77.54\%.}
    \label{size}
\end{figure*}

\section{Experiments}
\label{sec:exp}
\subsection{Dataset}
\vspace{-3pt}
Interactive Emotional Dyadic Motion Capture (IEMOCAP) \emph{et al.}\cite{busso2008iemocap} is the most widely used dataset in the SER field. It contains 12 hours of emotional speech performed by 10 actors from the Drama Department of University of Southern California. The performance is divided into two parts, improvised and scripted, according to whether the actors perform according to a fixed script. The utterances are labeled with 9 types of emotion-anger, happiness, excitement, sadness, frustration, fear, surprise, other and neutral state.

Due to the imbalances in the dataset, researchers usually choose the most common emotions, such as neutral state, happiness, sadness, and anger. Because excitement and happiness have a certain degree of similarity and there are too few happy utterances, researchers sometimes replace happiness with excitement or combine excitement and happiness to increase the amount of data \cite{li2018attention,zhao2019attention,neumann2019improving}. In addition, previous studies have shown that the accuracy of using improvised data is higher than that of scripted data,\cite{tarantino2019self,li2018attention} which can be due to the fact that actors pay more attention to expressing their emotion rather than the script during improvisation.

In this paper, following other published work, we use improvised data in the IEMOCAP dataset with four types of emotion--neutral state, excitement, sadness, and anger.
\vspace{-10pt}
\subsection{Evaluation Metrics}
We use weighted accuracy (WA) and unweighted accuracy (UA) for evaluation, which have been broadly employed in SER literature.
Considering that WA and UA may not reach the maximum in the same model, we calculate the average of WA and UA as the final evaluation criterion (indicated by \emph{ACC} below), \emph{i.e.}, we save the model with the largest average of WA and UA as the optimal model.
\subsection{Experimental Setup}
We randomly divide the dataset into a training set (80\% of data) and a test set (20\% of data) for 5-fold cross-validation. Each utterance is divided into 2-second segments, with 1 second (in training) or 1.6 seconds (in testing) overlap between segments. Although divided, the test is still based on the utterance, and the prediction results from the same utterance are averaged as the prediction result of the utterance.  Experience shows that a large overlap can make the recognition result of utterance more stable in testing.

\vspace{-10 pt}
\begin{figure}[htbp]
    \centering
    \includegraphics[width=0.65\columnwidth]{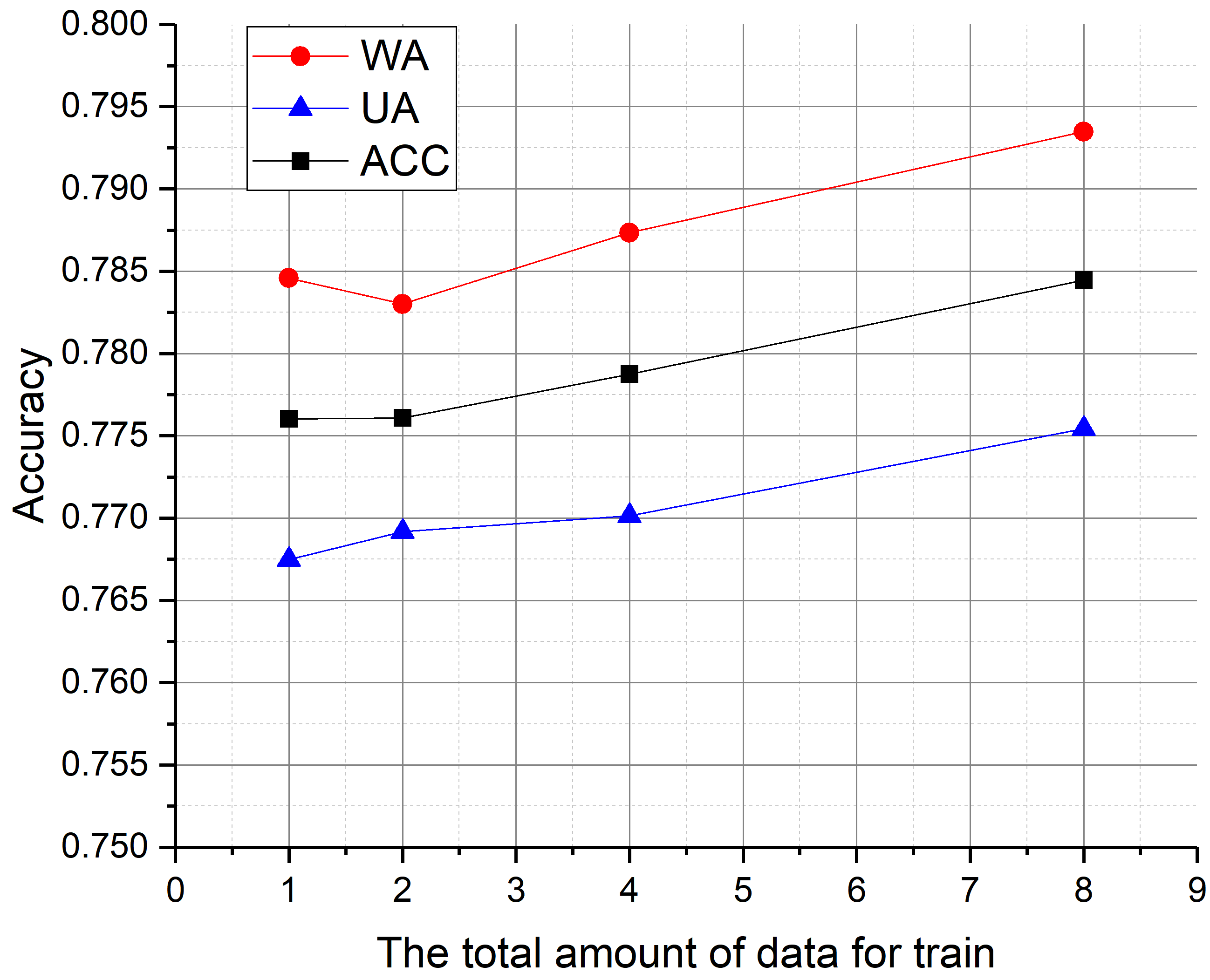}
    \caption{\textbf{Result of modifying the amount of data with VTLP} The horizontal axis refers to the total amount of data used for training. For example, 8 on the horizontal axis represents original data plus 7 replicas of augmented data. It can be seen that the more augmented data added the higher the accuracy. }
    \label{vtlp_amount}
\end{figure}

\begin{table*}[htbp]
\caption{Result of area feature selection}
\scriptsize
\centering
\subtable[\emph{WA}]{
\begin{tabular}{|l|l|l|l|}
\hline
\diagbox{\emph{Key}}{\emph{Value}}   & Max               & Mean     & Sum   \\ \hline
Max  & 0.7869         & 0.7880  & 0.7869 \\ \hline
Mean & 0.7864          & 0.7808 & 0.7846 \\ \hline
Sample  & \textbf{0.7911} & 0.7808 & 0.7851 \\ \hline
\end{tabular}
}
\qquad
\subtable[\emph{UA}]{        
\begin{tabular}{|l|l|l|l|}
\hline
\diagbox{\emph{Key}}{\emph{Value}}    & Max              & Mean     & Sum   \\ \hline
Max  & 0.7642          & 0.7627 & 0.7686 \\ \hline
Mean & 0.7639          & 0.7608 & 0.7675 \\ \hline
Sample  & \textbf{0.7705} & 0.7567  & 0.7657  \\ \hline
\end{tabular}
}
\qquad
\subtable[\emph{ACC}]{   
\begin{tabular}{|l|l|l|l|}
\hline
\diagbox{\emph{Key}}{\emph{Value}}    & Max              & Mean     & Sum   \\ \hline
Max    & 0.7755          & 0.7753 & 0.7777 \\ \hline
Mean   & 0.7751          & 0.7708  & 0.7760 \\ \hline
Sample & \textbf{0.7808} & 0.7687 & 0.7753 \\ \hline
\end{tabular}
}

\label{areafeature} 
\end{table*}

\begin{figure}[htb]
    \centering
    \subfigure[Original LogMel]{\includegraphics[width=1.1 in]{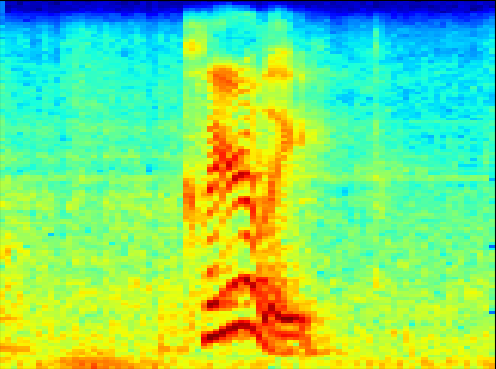}}
    \subfigure[CNN]{\includegraphics[width=1.1 in]{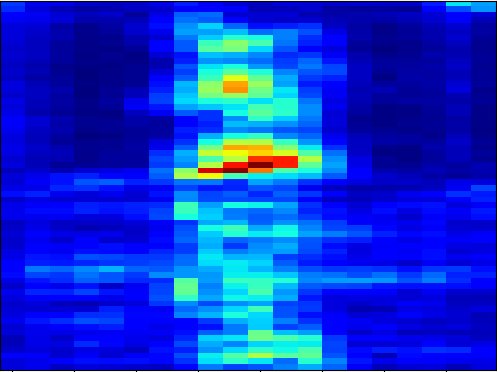}}
    \subfigure[Area attention]{\includegraphics[width=1.1 in]{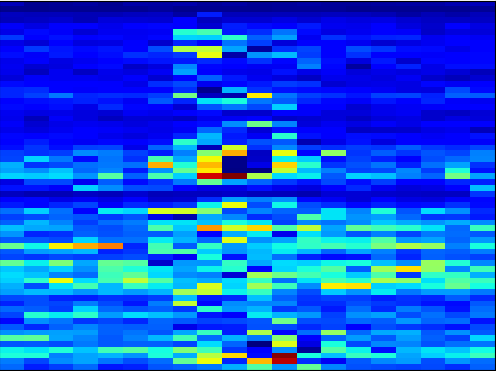}}
    \caption{\textbf{Feature representation} CNN attends more on areas with high-energy or sharp energy changes. This benefits the sudden and intense emotions. The area attention model not only attends on these areas but also extends that to the time domain (horizontal axis), which enables it to distinguish those long-term emotions.}
    \label{visual}
\end{figure}

\vspace{-10pt}
\begin{table}[htbp]
\centering
\caption{Ablation experiment}
\footnotesize
\begin{tabular}{|l|l|l|l|}
\hline
\textit{\textbf{Model}}  & \textit{\textbf{WA}} & \textit{\textbf{UA}} & \textit{\textbf{ACC}} \\ \hline
CNN                      & 0.7467               & 0.7222               & 0.7345                \\ \hline
CNN+VTLP                 & 0.7891               & 0.7683               & 0.7787                \\ \hline
Attention                & 0.7807               & 0.7628               & 0.7718                \\ \hline
Attention+VTLP           & 0.7879               & 0.7734               & 0.7807                \\ \hline
Area attention      & 0.7911               & 0.7705               & 0.7808                \\ \hline
Area attention+VTLP & \textbf{0.7934}      & \textbf{0.7754}      & \textbf{0.7844}       \\ \hline
\end{tabular}
\label{ablation}
\end{table}

\vspace{-10pt}
\subsection{Experimental Results}

\noindent\textbf{Selection of maximum area size} Optimal maximum area size is investigated on both the original data and the augmented data with VTLP, respectively. The result is shown in Fig \ref{size}. It can be seen that when trained on the original data set, the model with the max area size of 4x4 achieved the highest \emph{ACC}, followed by 3x3. When trained on the augmented data set, the model with the max area size of 3x3 achieved the highest \emph{ACC}. In most cases, the use of enhanced data brings an accuracy increase of more than 0.5\% absolute. Therefore, we suggest using a max area size of 3x3 and using VTLP for data augmentation.

\noindent\textbf{Selection of area features} Experiments are conducted to investigate the performance of using various area features. For \emph{Key}, we selected \emph{Max, Mean} and \emph{Sample}; for \emph{Value}, we selected \emph{Max, Mean} and \emph{Sum}.The \emph{Sample} refers to adding a perturbation proportional to the standard deviation on the basis of \emph{Mean} when training, which is calculated according to Eq.\ref{sample} where $x$ is a sample and $\mu$ and $\sigma$ are the mean and standard variance, respectively. $\xi$ is a random variable assuming $\mathcal{N}(0,1)$ distribution.  We use \emph{K-V} to represent the model selected \emph{K} as \emph{Key} and \emph{V} as \emph{Value}. 
\begin{equation}
    \begin{split}
    &  x = \mu + \sigma *\xi, \ \  \ \ \text{where} \ \xi \sim  \mathcal N(0,1)
    \end{split}
    \label{sample}
\end{equation}
Table \ref{areafeature} shows the result. It can be observed that the \emph{Sample-Max} achieved the highest \emph{ACC} and the \emph{Sample-Mean} achieved the lowest \emph{ACC}. There is little difference in \emph{ACC} in other cases. We speculate that it is because perturbed \emph{Key} in training introduces greater randomness.

\noindent\textbf{Amount of augmented data} Experiments are also carried out to study the impact of amount of augmented data under VTLP to the SER performance, as demonstrated in Fig \ref{vtlp_amount}.  It can be observed that with more replicas of augmented data added in the training, the accuracy increases.

\noindent\textbf{Ablation study} We conducted ablation experiments on the model without the attention layer (only CNN) and the model with an original attention layer (equivalent to 1x1 max area size). Table \ref{ablation} shows the result. It can be seen that the area attention and VTLP enables the model to achieve the highest accuracy. As a case study, we visualize the feature representations of an input logMel right before the fully connected layer of the learned model in Fig \ref{visual}. It clearly shows that compared to the conventional CNN with more localized representation, area attention tends to cover a wide context along the time axis, which is one of the reasons that area attention can outperform CNN. Also from Table \ref{ablation} we can see that when the model becomes stronger, the improvement brought by VTLP marginally decreases. This is because VTLP conducts label-preserving perturbation to improve the robustness of the classifier. When the model gets stronger with attention or multiscale area attention, the model itself becomes more robust, which may offset to certain degree the impact of VTLP.   
\vspace{-10pt}
\begin{table}[htbp]
\centering
\footnotesize
\caption{Accuracy comparison with existing SER results}
\begin{tabular}{|l|l|l|l|}
\hline
\textit{\textbf{Method}}                                         & \textit{\textbf{WA(\%)}}    & \textit{\textbf{UA(\%)}}    & \textit{\textbf{Year}} \\ \hline
Attention pooling (P. Li \emph{et al.})             \cite{li2018attention}             & 71.75 & 68.06 & 2018 \\ \hline
CTC + Attention (Z. Zhao \emph{et al.})  \cite{zhao2019attention}                         & 67.00    & 69.00    & 2019 \\ \hline
Self attention (L. Tarantino \emph{et al.})  \cite{tarantino2019self}                     & 70.17 & 70.85 & 2019 \\ \hline
BiGRU (Y. Xu \emph{et al.})    \cite{xu2020hgfm}                    & 66.60  & 70.50  & 2020 \\ \hline
\begin{tabular}[c]{@{}l@{}}Multitask learning + Attention\\ (A. Nediyanchath \emph{et al.})\cite{nediyanchath2020multi}\end{tabular} & 76.40  & 70.10  & 2020 \\ \hline
Head fusion (Ours) \cite{xu2020improve}                             & 76.18 & 76.36 & 2020 \\ \hline
Area attention (Ours)                           & \textbf{79.34}  & \textbf{77.54}  & 2020 \\ \hline
\end{tabular}
\label{compare}
\end{table}

\noindent\textbf{Comparison with existing results}  As shown in Table \ref{compare}, we compared our accuracy with other published SER results in recent years. These results use the same data set and the evaluation metrics as our experiment.

\section{Conclusion}
\label{sec:conclusion}
In this paper, we have applyed multiscale area attention to SER and designed an attention-based convolutional neural network, conducted experiments on the IEMOCAP data set with VTLP augmentation, and obtained 79.34\% WA and 77.54\% UA. The result is state-of-the-art.

In future research, we will continue to work along the lines by improving the application of attention on SER and apply more data augmentation methods to SER.

{
\small
\bibliographystyle{IEEEbib}
\bibliography{reference}
}

\end{document}